
\def   \ni {\noindent}
\def   \cl {\centerline}

\def   \ssk {\vskip  5truept}

\def   \bsk {\vskip 15truept}

\def   \newline {\hfil\break}

\input psfig.tex
\magnification=1000
\hsize 5truein
\vsize 8truein
\font\abstract=cmr8

\font\text=cmr10
\font\affiliation=cmssi10
\font\author=cmss10

\font\title=cmssbx10 scaled\magstep2

\def\ref{\par\noindent\hangindent 15pt}
\nopagenumbers
\null
\vskip 3.0truecm
\baselineskip = 12pt

{\title
Evolution of a Magnetic Flux Tube in a Sunspot Penumbra
\ni
}                                            
\bsk \bsk
{\author
 K. Jahn$^{1,2}$, R. Schlichenmaier$^{3}$, H.U. Schmidt$^{1}$
\ni
}                                            
\bsk
{\affiliation
$^1$Max Planck Institut f\"ur Astrophysik, \hfill\break
Karl Schwarzschild Str.1, D-85748 Garching, Germany

$^2$Astronomical Observatory of the Warsaw University,  \hfill\break
Al. Ujazdowskie 4, 00-478 Warsaw,
Poland, (e-mail: crj@sirius.astrouw.edu.pl)

$^3$Max Planck Institut f\"ur Extraterrestrische Physik, \hfill\break
Karl Schwarzschild Str.1, D-85748 Garching, Germany.
}                                            
\bsk
\cl {\it (Received ............. )}
\bsk
\baselineskip = 9pt
{\abstract
\ni
The motion  of an individual magnetic flux tube inside the penumbra of
a sunspot is studied numerically.
Here, we present preliminary results.
The thin flux tube approximation together with a simplified radiative
heat exchange with the surroundings is used to study the evolution of
a flux tube embedded into a background given by a global
magneto-static sunspot model.
The investigation is undertaken in order to verify the conjecture that
convection in sunspot penumbrae occurs by an interchange of magnetic
flux tubes.
The code being developed can be used to study dynamic aspects of
filamentary structure in the penumbra:
the temporal and spatial fluctuations of the temperature and the
magnetic field, the motion of bright penumbral grains, or the Evershed
effect.
Here we present the evolution of a wave formed by the tube whose
fragment emerges in the penumbral photosphere and migrates towards the
umbra.
The properties of this wave show qualitative features of the observed
bright penumbral grains with corresponding upward velocity and its
correlation with brightness and the inclination of the magnetic field,
and also of the Evershed effect.
}                              	               
\bsk

\baselineskip = 12pt
{\text                                         
\ni 1. INTRODUCTION

High resolution observations of filamentary sunspot penumbrae give
some detailed hints at a particular type of convection that seems to
occur there (see an extensive discussion in Jahn \& Schmidt, 1994).
In summary, the inclination of the magnetic field exhibits substantial
azimuthal fluctuations which preserve their radial coherence and are
well correlated with the brightness of filaments (Title et al., 1993):
the dark ones are permeated by an almost horizontal field whereas the
bright ones contain a more vertical field.
The velocity field shows similar correlations: the horizontal
Evershed effect occurs in the dark regions and upward flow in brighter
ones.
Additionally, an inward motion of bright penumbral grains towards the
umbra is observed (Muller, 1992), which may be associated with the
inward part of a balanced two-way migration of the magnetic flux.
Those features observed in the penumbral photosphere concur with the
concept of an interchange convection in the penumbra which appears to
have a deep structure (Schmidt 1987, Jahn 1989, Solanki \& Schmidt 1993).
This hypothetical convection would be accomplished by a non-local
exchange of azimuthally narrow but radially elongated magnetic flux
tubes that extend many scale heights below the photosphere (Schmidt
1991, Jahn 1992).
It would set in when the total flux and with it the inclination of the
field lines at the edge of a magnetic flux concentration exceeds some
critical value.
Rucklidge et al. (1995) studied an abrupt development of the penumbra
with a simple model based on this concept.
Some anticipated properties of interchanging flux tubes have been used
to approximate the average structure of a deep penumbra by a one
dimensional stratification of the matter, and to construct a
tripartite magneto-static model for a sunspot (Jahn \& Schmidt 1994).
Now we intend to construct a 2-D model for the penumbra based on a
more detailed study of the dynamics of the interchange convection.
This might be achieved in an iterative manner by following the
non-linear time-dependent evolution of a single magnetic flux tube
embedded into a background which asymptotically assumes the
thermodynamic and magnetic properties determined by the properties of
the flux tube itself averaged over all phases of its motion.

In this progress report we present preliminary results of a study of
the migration of a flux tube embedded into the fixed pregiven penumbra
of the tripartite model, and compare its dynamic properties with the
observed evolution of bright penumbral grains.

\midinsert
\vskip 0.5truecm
\par\noindent
\overfullrule 0pt
\hskip -1truecm
\centerline{\psfig{figure=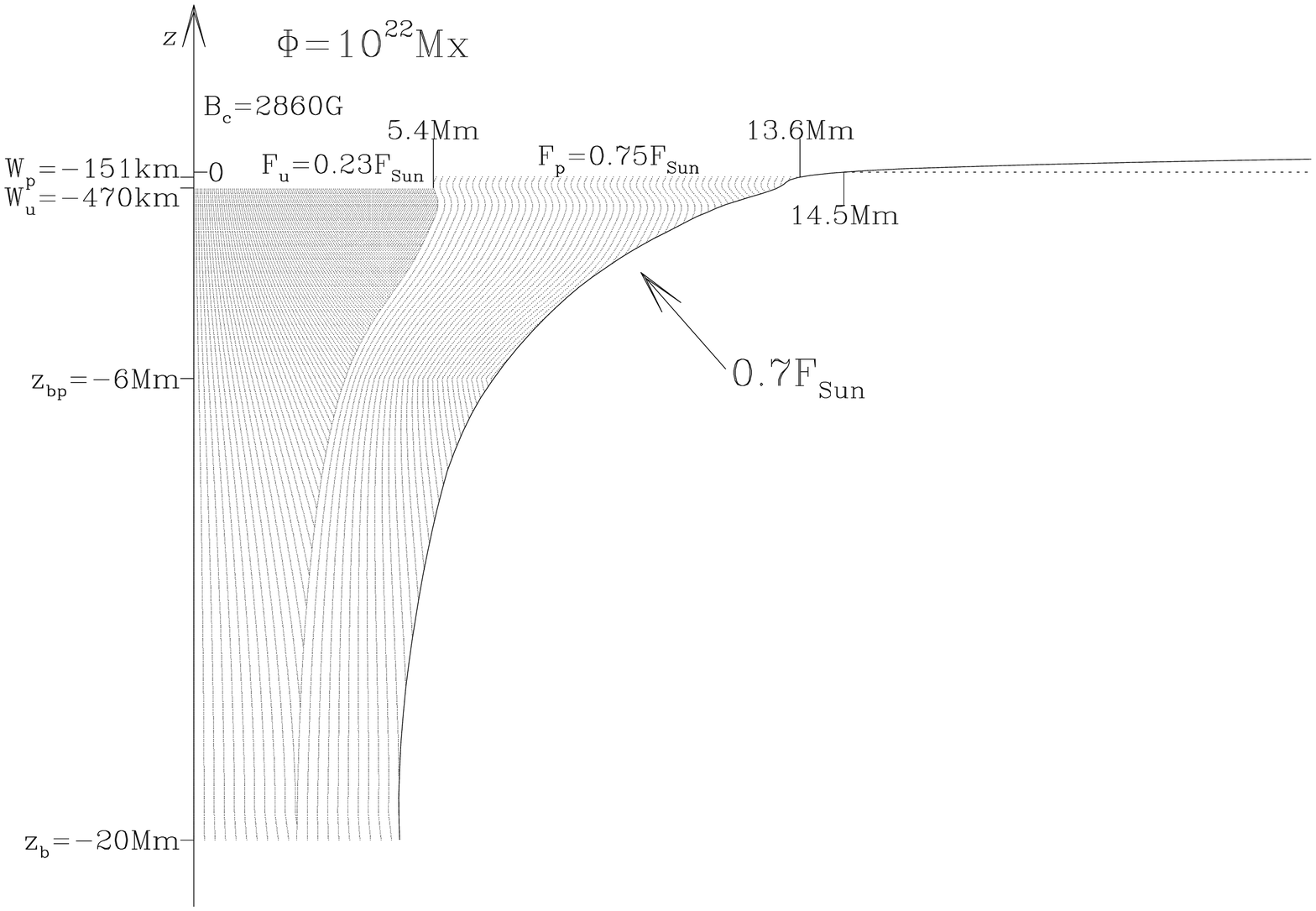,height=9truecm,width=14truecm}}
\leftskip=1truecm
\rightskip=1truecm
\noindent
{\abstract FIGURE 1.
Parameters of the tripartite sunspot model (cf. Jahn \& Schmidt 1994) }
\vskip 0.5truecm
\leftskip=0truecm
\rightskip=0truecm
\endinsert

\ni 2. THE MODEL

The structure of the magnetic flux element has been approximated by a
1-D thin flux tube, and the Lagrangian formalism has been adapted to
describe the 2-D dynamics of the tube.
The equations governing the time evolution of the tube are basically
the same as those used by Moreno-Insertis (1986) except that we allow
for the radiative heat exchange between the tube and the surroundings
and calculate the rate of the entropy change due to radiation.
This is made in a similar manner as by Montesinos \& Thomas (1993)
who distinguished between large optical depths ($\tau>2/3$), where the
entropy change is proportional to the temperature difference between
the tube and the surroundings (according to Spiegel's (1957)
analysis), and small optical depths ($\tau<2/3$), where this change is
proportional to the difference in the local values of the Planck
function (Kalkofen \& Ulmschneider 1977).

Another difference is related to the 2-D nature of the force-free
magnetic field of the background in which our tube is embedded.
Although the stratification of the gas in the tripartite model is one
dimensional, the magnetic pressure has a lateral gradient, so that the
distribution of the total pressure (gas plus magnetic) is
two-dimensional.
This leads to additional terms in the equation of motion and in the
condition of mechanical equilibrium of the tube with the surroundings.
Finally, the set of equations describing the tube's evolution is
completed by the same equation of state as that used in the
tripartite model, i.e. with partial ionization effects included for
a gas of solar chemical composition.

All parameters of the particular tripartite model used as background
in the present simulations are given in Fig.~1 (see also Fig.1 in Jahn
\& Schmidt, 1994).
\bsk

\ni 3. RESULTS

A thin tube with a magnetic flux equal to 10$^{15}$ Mx is originally
placed on the inner side of the magnetopause of the tripartite model.
The corresponding diameter of the tube at the penumbral photosphere is
equal to about 10~km.
Such a tube is much too small to represent directly any of the
fine-structure elements observed in the penumbra, but our simplified
treatment of the radiative transfer does not allow to consider much
larger flux tubes.
Otherwise, we would have to take into account an internal radiative
diffusion.
The initial stratification  along the tube is identical to that in
the penumbra.
The end points of the tube have been placed far from the photosphere
of the penumbra: the lower one at a depth of 15~Mm and the upper one
on a sphere 25~Mm away from the center of the spot.
For the time being both points are kept at fixed positions on the
magnetopause during the evolution of the tube.

When the tube is slightly perturbed locally, its near photospheric
portion at depths less than about 500~km, starts to oscillate across
the magnetopause, because of two competing forces: the buoyancy which
expels the tube from the more dense stratification of the quiet sun,
and the curvature force acting in the opposite direction.
But each time the tube intrudes into the quiet sun it heats up by
radiative exchange which is still efficient at those depth in a
nonlinear fashion, so that the amplitude of the oscillation grows with
each cycle.
Eventually, the tube gains a buoyancy large enough to overcome the
curvature force, then emerges in the penumbral photosphere and migrates
further towards the umbra.

\midinsert
\vskip 0.5truecm
\par\noindent
\overfullrule 0pt
\hskip -1truecm
\centerline{\psfig{figure=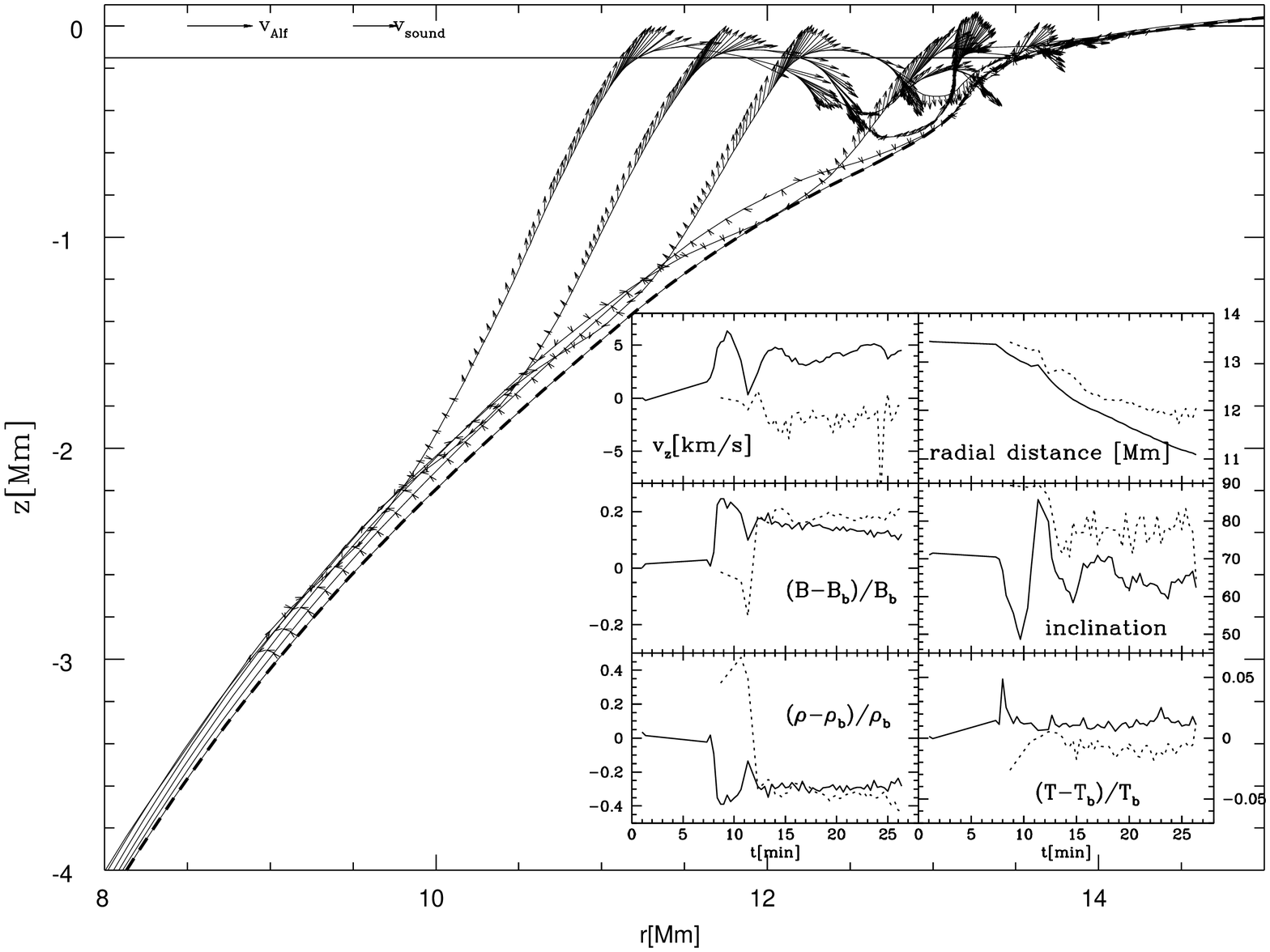,height=11truecm,width=15truecm}}
\leftskip=1truecm
\rightskip=1truecm
\noindent
{\abstract FIGURE 2.
Migration of a flux tube in the penumbra plotted with a time step of 5
minutes.
Dashed line corresponds to the magnetopause, and arrows show the flow
of the matter.
Insertions show the time history of the three points crossing the
photosphere: the emerging (continuous line) and sinking one (dotted
line).
Shown are: fluctuations of the density, temperature, and the
magnetic field relatively to the corresponding background values,
the inclination of the tube to the vertical direction,  vertical
component of the velocity of plasma in the tube, and the distance
of the points from the spot's center
}
\vskip 0.5truecm
\leftskip=0truecm
\rightskip=0truecm
\endinsert

Figure~2 shows the geometry of the outer part of the penumbra and the
shape of the tube at sample times differing form each other by five
minutes.
The arrows along the tube indicate the velocity of the matter.
The length of arrows corresponding to the Alfv\`en and sound speed at
the photosphere are also marked in the figure.
Initially the tube crosses the penumbral photosphere only once, but as
it proceeds inwards its part laying above the photosphere near the
edge of the spot sinks, so that the tube assumes a sine-wave shape
with three intersections with the photosphere.
The time history of different parameters of the two inner
intersection points are plotted in the insertions, where continuous
and dotted lines correspond to the ascending and the descending
intersection respectively.
The third intersection retracted after about 10 minutes and never
intruded further than 500~km into the penumbra.
For a short time two more intersections appeared near the penumbral
boundary, because the tube continuously oscillates across the
magnetopause in this region.
The last shape of the tube shown in the Fig.~2 may suggest that most
probably there will soon be generated a second wave crossing the
surface and traveling inward.
However, problems with our oversimplified lower boundary condition
do not allow to continue the run.
Therefore, we concentrate at present on the features of the first wave
which traverses about 2.5~Mm in the penumbra during the simulation.
It forms a flattened arch above the photosphere with a typical height
of about 50~km which has a length that slowly increases with time from
about 300 to 800~km (see insertions in  Fig.~2).

The flow of the gas in the arch is directed opposite to the migration
of the tube.
It assumes speeds close to sonic, or even supersonic in the
descending intersection.
It should be noted that our code does not yet allow a proper treatment
of shocks.
This streaming along the arch has a nature similar to a siphon flow
(Meyer \& Schmidt, 1968).
The velocity of this flow in our model, of about 7~km/s, is larger
than the observed maximum velocity of the Evershed effect, i.e.
6~km/s, reached at the outer penumbra boundary (see e.g. Muller, 1992).

Properties of both intersection points show some qualitative
similarities to those observed in penumbral fine-structure, i.e. flow
velocities, magnetic field directions, and their correlations with
brightness.
Thus, the ascending intersection has a more vertical magnetic field
than the mean penumbral field in the model.
The matter flows upward there and is hotter than the ambient medium.
That concurs with the upward velocity of the bright matter measured in
penumbrae (Beckers 1981, Degenhardt \& Wiehr 1991).
The other intersection is more horizontal, but shows a downflow with a
substantial vertical component of the velocity as the gas is cooler
than the surrounding background in the photosphere of the penumbra.

It is tempting to identify the ascending intersection of the arch with
a bright penumbral grain, despite the fact that it does not possess
all the quantitative features of the grains.
It moves inward with a speed of about 2~km/s, the descending one
somewhat slower.
These values are by a factor of four larger than the observed
velocities of the penumbral grains (Muller, 1992).
The large velocity may be ascribed to the mechanism of heating of the
flux tube in our present model.
It is achieved by the radiative exchange whenever the tube intrudes
into the quiet sun.
The radiative time scale increases abruptly with depth below the
photosphere due to increasing opacity.
Thus, the tube can acquire the energy efficiently only over a limited
range of depths, so that the buoyancy can be increased in a short
fragment of the tube only, which then rises thereby forming a
relatively short and fast-moving wave.
One might expect that more efficient heating at larger depths might be
achieved by complex magneto-convective processes.
This then would make the tube buoyant over larger lengths, and thereby
would induce longer and slower waves.
\bsk

\ni {4. CONCLUSIONS}

We have developed a code that allows investigations of the motion of a
thin flux tube embedded in a pregiven magnetic field and background
stratification extending over many scale heights in the solar
convection zone and atmosphere.
Preliminary simulations presented here show the effect of radiative
heating of the flux tube immediately outside the magnetopause occurring
over a relatively small range of depths.
It induces a wave-like inward migration of the tube, and the
properties of the migrating wave correspond in general to the observed
features of the fine structure in the penumbra, although the
particular values differ by a factor of a few from the observed
quantities.
Probably, a better agreement can be achieved with a more efficient
heating of the tube at larger depths.
That requires, however, a model for the heat exchange across an
inclined magnetopause in convectively unstable stratification,
like that of Schmidt et al. (1986).
Also a correct treatment of shocks should be included in the code,
because supersonic velocities along the tube appear in the
photospheric layers.
\ssk

{\bf Acknowledgments}. We acknowledge the support of the NATO
Advanced Study Institute. KJ acknowledges the hospitality and the
support from the Max Planck Institut f\"ur Astrophysik in Garching.
RS likes to thank Prof. G. Haerendel for providing the possibility to
work on this subject.
This work was also supported by the KBN grant 2 P304 004 007.
\ssk

\vskip 0.1truecm
\ni {REFERENCES}
\ssk

\ref Beckers J.M, 1981, in `` The Sun as a Star'', S. Jordan (ed.),
NASA-SP-450, Washington, p.11

\ref Degenhardt D., Wiehr E., 1991, A\&A {\bf 252}, 821

\ref Jahn K., 1989, A\&A {\bf 222}, 264

\ref Jahn K., 1992, in ``Sunspots: Theory and observations'', J.H. Thomas
\& N.O. Weiss (eds.), Kluwer, Dordrecht, p.139

\ref Jahn K., Schmidt H.U., 1994, A\&A {\bf 290}, 295

\ref Kalkofen W., Ulmschneider P., 1977, A\&A {\bf 57}, 193

\ref Meyer F., Schmidt H.U., 1968, Zeitschrift f. Angew. Mathematik
u. Mechanik {\bf 48}, 218

\ref Montesinos B., Thomas J.H., 1993, ApJ {\bf 402}, 314

\ref Moreno-Insertis F., 1986, A\&A {\bf 166}, 291

\ref Muller R., 1992, in ``Sunspots: Theory and observations'', J.H. Thomas
\& N.O. Weiss (eds.), Kluwer, Dordrecht, p.175

\ref Rucklidge A.M., Schmidt H.U., Weiss N.O., 1995, MNRAS {\bf 273}, 491

\ref Schmidt H.U., 1987, in ``The role of fine-scale magnetic field on the
structure of the solar atmosphere'', E.H. Schr\"oter, M. Vazquez,
A.A. Wyller (eds.), Cambridge University Press, p.219

\ref Schmidt, H.U.: 1991, Geophys. Astrophys. Fluid Dyn. {\bf 62}, 249

\ref Schmidt H.U., Spruit H.C., Weiss N.O., 1986, A\&A {\bf 158}, 351

\ref Solanki S.K., Schmidt H.U., 1993, A\&A {\bf 267}, 287

\ref Spiegel E.A., 1957, ApJ {\bf 126}, 202

\ref Title A.M., Frank Z.A., Shine R.A. et al., 1993, ApJ {\bf 403}, 780

}                                                 

\end